\documentclass[journal]{IEEEtran}

\usepackage{color}

\usepackage{cite}

\ifCLASSINFOpdf
  \usepackage[pdftex]{graphicx}
\else
\fi

\usepackage{amsmath}
\usepackage{amsfonts}

\usepackage{algorithmic}

\usepackage{multirow}

\begin{document}

\title{Intelligent Parameter Tuning in Optimization-based Iterative CT Reconstruction via Deep Reinforcement Learning}

\author{Chenyang Shen, Yesenia Gonzalez, Liyuan Chen, Steve B. Jiang, Xun Jia*
\thanks{Chenyang Shen (Chenyang.Shen@UTSouthwestern.edu), Yesenia Gonzalez (Yesenia.Gonzalez@UTSouthwestern.edu), Liyuan Chen (Liyuan.Chen@UTSouthwestern.edu), Steve B. Jiang (Steve.Jiang@UTSouthwestern.edu), and Xun Jia (Xun.Jia@UTSouthwestern.edu) are with the Department of Radiation Oncology, University of Texas Southwestern Medical Center, Dallas, TX 75390 USA. Asteroid indicates the corresponding author.}}

\markboth{Intelligent Parameter Tuning in Iterative CT Reconstruction}{}

\maketitle


\begin{abstract}
A number of image-processing problems can be formulated as optimization problems. The objective function typically contains several terms specifically designed for different purposes. Parameters in front of these terms are used to control the relative weights among them. It is of critical importance to tune these parameters, as quality of the solution depends on their values. Tuning parameter is a relatively straightforward task for a human, as one can intelligently determine the direction of parameter adjustment based on the solution quality. Yet manual parameter tuning is not only tedious in many cases, but becomes impractical when a number of parameters exist in a problem. Aiming at solving this problem, this paper proposes an approach that employs deep reinforcement learning to train a system that can automatically adjust parameters in a human-like manner. We demonstrate our idea in an example problem of optimization-based iterative CT reconstruction with a pixel-wise total-variation regularization term. We set up a parameter tuning policy network (PTPN), which maps an CT image patch to an output that specifies the direction and amplitude by which the parameter at the patch center is adjusted. We train the PTPN via an end-to-end reinforcement learning procedure. We demonstrate that under the guidance of the trained PTPN for parameter tuning at each pixel, reconstructed CT images attain quality similar or better than in those reconstructed with manually tuned parameters. 
\end{abstract}

\begin{IEEEkeywords}
Image reconstruction - iterative methods, Machine learning, Inverse methods, x-ray imaging and computed tomography
\end{IEEEkeywords}

\ifCLASSOPTIONpeerreview
\fi
\IEEEpeerreviewmaketitle

\section{Introduction}
%
%
%
%

\IEEEPARstart{A}{} number of medical image-processing problems can be formulated as solving optimization problems. In such problems, the objective function typically contain several terms carefully designed for different purposes. A set of parameters are used to control the relative weights of these terms in order to achieve a satisfactory solution quality. Take a typical problem of iterative Computed Tomography (CT) reconstruction as an example, it can be formulated as 
\begin{equation}
\label{Eq:CTreconGeneral}
f^* = \arg\min_f\frac{1}{2}|Pf-g|^2+\lambda~R[f],
\end{equation}
where $f^*$ is the image vector to be reconstructed by solving the optimization problem, $P$ stands for the x-ray projection operator, and $g$ the measured projection data. The first term is a data-fidelity term, minimizing of which ensures agreement between $f^*$ and the measurement $g$. $R[f]$ stands for a regularization term specifically designed to enforce quality of the solution image from a certain aspect, e.g. piece-wise smoothness. $\lambda$ is the parameter that is used to control the trade-off between this regularization term and the data-fidelity term. Over the years, a number of regularization terms have been developed to successfully restore a solution $f^*$ using undersampled or noisy measurement $g$. Examples include, but are not limited to, total variation (TV)\cite{Rudin:PhyD:1992,Sidky:PMB:2008,Jia:MP:2010}, tight frame (TF)\cite{Dong:IAS:2010,Jia:PMB:2011}, and nonlocal means (NLM)\cite{Lou:JSC:2010,Chen:PhyMed:2016,Jia:MP:2012}.

Despite the success, parameter tuning in these optimization-based image processing problems is inevitable. Manually adjusting the parameters for the best image quality is not uncommon in literature\cite{Sidky:PMB:2008,Jia:MP:2010,Jia:PMB:2011,Chen:PhyMed:2016,Jia:MP:2012}. Yet this is a tedious approach, as one has to carefully navigate through the parameter space to find the optimal value. The required efforts and human time impede clinical applications of those novel image-processing methods. Moreover, manual parameter tuning becomes an increasingly challenging task in those problems with multiple regularization terms. An extreme example is CT reconstruction but with weighting parameter freely adjustable at each pixel\cite{Guo:SPIE:2010,Tian:PMB:2011}. Clearly, the substantial amount of parameters makes manual parameter tuning infeasible. Therefore, it is highly desirable to develop a method for automatic parameter tuning. Over the years, this problem has attracted a lot of research interests. For instance, generalized cross validation and L-curve methods have been used to choose the regularization parameter\cite{Golub:Techno:1979,Hansen:SIAMreview:1992,Ramani:TIP:2012}. It has also been proposed to develop a method to assess image quality, which can be used to guide parameter adjustment towards the direction of improving the quality\cite{Zhu:TIB:2010, Liang:TIP:2016}. In certain contexts, such as the CT reconstruction problem, it may be even possible to estimate the level of data contamination based on physics or mathematical principles. This can provide valuable information to set the parameter values\cite{Bai:JxST:2017}. Despite these efforts, a practical solution that is  applicable to general problems still does not exist, calling for further investigations. 

Although it is quite difficult for a computer to automate the parameter tuning process, this task seems to be less of a problem for humans. One typically has a strong intuition about which direction the parameter should be adjusted based on the observed image quality. Again, let us take the iterative CT reconstruction problem in Eq.~(\ref{Eq:CTreconGeneral}) as an example. By looking at the solution image, one knows that the regularization term needs to be enhanced, if the solution appears to be noisy, or be relaxed otherwise. Based on this fact, it is of interest and importance to model this remarkable intuition and capability in an intelligence system, which can then be used to solve the parameter tuning problem from a new angle. 

Not until recently does the tremendous success in deep-learning regime shine a light in this direction. In the past a few years, deep learning has clearly demonstrated its power in numerous medical image processing problems\cite{Han:CoRR:2016,Kang:CoRR:2016,Li:Fully3d:2017,Chen:BOE:2017,Chen:TMI:2017,Cheng:Fully3d:2017}. More importantly, it was found that human-level intelligence can be spontaneously generated via deep-learning schemes, which enables a system to perform a certain task in a human-like fashion, or even better than humans. In a pioneer work, an artificial intelligent system was developed to realize human-level control of Atari computer games\cite{Mnih:Arxiv:2013,Mnih:Nature:2015}. Employing a deep Q-network approach, the system was trained through the framework of deep reinforcement learning to learn how to interact with the environment, i.e. play an Atari game. The results were remarkable: the trained system was able to achieve a level comparable to that of a professional human players in a set of 49 Atari games. 

Motivated by this fact, we propose in this paper to develop an intelligent system to accomplish the parameter tuning task in optimization-based image-processing problems. We take a CT reconstruction problem as a example to demonstrate our idea. Specifically, we will develop a parameter tuning policy network (PTPN), which can intellectually determine the direction and magnitude of parameter tuning by observing an input image patch. The rest of this paper is organized as follows. Sec. II will introduce the example problem of TV-based CT reconstruction with pixel-wise regularization. We will also describe the PTPN structure and how to train it to develop the skill of parameter tuning. Sec. III will present our validation studies and results. Finally, we will make some discussions in Sec. IV and conclude the study in Sec. V.

\section{Methods}

\subsection{An example CT reconstruction problem}
In this paper, we consider the following iterative CT reconstruction problem as an example to demonstrate our idea:
\begin{equation}
\label{Eq:CTrecon}
f^* = \arg\min_f\frac{1}{2}|Pf-g|^2+|\lambda\cdot\nabla f|.
\end{equation}
This approach falls in to the regime of TV-based regularization\cite{Rudin:PhyD:1992}, which penalizes the $L_1$ norm of the image gradient to ensure image smoothness while preserving edges. In the second term of the objective function, we consider a general case that extends $\lambda$ into a vector. Each entry of $\lambda$ controls the weight of an image pixel. The substantially higher amount of parameters in this example problem than a typical single-parameter TV model highlight the need for an automatic parameter tuning system.

There are a number of novel numerical algorithms to solve this optimization problem \emph{for fixed parameter $\lambda$}\cite{Goldstein:SIAM:2009,Chambolle:JMIV:2011,Boyd:FTML:2011}. In this study, we use the alternating direction method of multipliers (ADMM)\cite{Boyd:FTML:2011}. It introduces an auxiliary variable $d$ and adds a constraint $d=\nabla f$. The problem can be handled by tackling the augmented Lagrangian:
\begin{equation}
\begin{split}
\mathcal{L}(f,d,\Gamma) &= \frac{1}{2}|Pf-g|^2+|\lambda\cdot d| +\frac{\beta}{2}|\nabla f-d|^2\\
& + \langle\Gamma,\nabla f-d\rangle,
\end{split}
\end{equation}
where $\beta$ is a parameter in the algorithm. Major steps of the ADMM algorithm is outlined in Fig.~\ref{Fig:ADMM}. Due to the large scale of the reconstruction probelm, the matrix inverse operation in Line 2 is achieved using conjugate gradient algorithm\cite{Golub:2013:matrix}. 

\begin{figure}
\begin{algorithmic}[1]
\REQUIRE $f^0$, $d^0, \Gamma^0$, $\beta$, stopping criteria $\delta$, $i = 0$.
\FOR{$i=1,2,\ldots$ }
	\STATE{$f^{i+1}=(P^TP-\beta\Delta)^{-1}(P^Tg-\beta\nabla d^i+\nabla 	\Gamma^i)$;}
	\STATE{$d^{i+1}=\mathrm{shrinkage}_{\lambda/\beta}(\nabla f^{i+1}+\frac{\Gamma^i}{\beta})$;}
	\STATE{$\Gamma^{i+1}=\Gamma^i+\beta(\nabla f^{i+1}-d^{i+1})$;}
	\IF{$|f^{i+1}-f^i|/|f^i|\leq \delta$}
		\STATE{Stop};
	\ENDIF
\ENDFOR
\end{algorithmic}
\caption{ADMM algorithm used to solve the problem in Eq.~\ref{Eq:CTrecon}.}
\label{Fig:ADMM}
\end{figure}

\subsection{System setup}

Our system tunes parameters $\lambda(x)$ in an iterative manner. Specifically, at the iteration step $k$, it observes the result generated by the image reconstruction system using ADMM algorithm at its convergence, $f^k$. Note that here $f^k$ is the solution at the convergence of the ADMM algorithm, rather than the image during the ADMM iteration. For each pixel $x$, the image patch centering around this pixel, denoted as $S_{f^k}(x)$ is fed to the parameter tuning system. The system then outputs direction and magnitude by which the parameter $\lambda^k(x)$ is adjusted. Here, we explicitly associate $\lambda(x)$ with the index $k$, as it will vary from step to step. Such a process continues, until a stopping criteria is met. 

We would like to achieve the parameter tuning capability using the optimal action-value function in the Q-learning regime\cite{Watkins:ML:1992}. This function is defined as 
\begin{equation}
\begin{split}
	Q^*&(s, a)\\
	&=\max_{\pi}[r^k+\gamma r^{k+1}+\gamma^2 r^{k+2}+\cdots|s_k =s,a_k=a,\pi ],
\end{split}
\end{equation}
where $r^k$ is the reward at iteration step $k$, $\gamma\le 1$ is a discount factor, and $\pi$ stands for the parameter tuning policy: taking an action $a$ after observing a state $s$. Here, we consider a deterministic policy that generates a unique action $a$ based on the observed state $s$. Specifically, we follow a greedy strategy that selects the action maximizing the $Q^*$ value under the input $s$, i.e. $a= \arg\max_{a'}~Q^*(s,a')$. In the particular problem of interest here, the state $s$ is an image patch $S_{f^k}(x)$. We consider five possible output actions: keeping the parameter $\lambda^k(x)$ unchanged, increasing or decreasing it by $10\%$, and increasing or decreasing it by $50\%$. We choose the values of $50\%$ or $10\%$ as possible amounts of changes in the system, as we expect these values will not critically affect the capability of parameter tuning of our system. 

\begin{figure*}[htbp]
  \includegraphics[width=\textwidth]{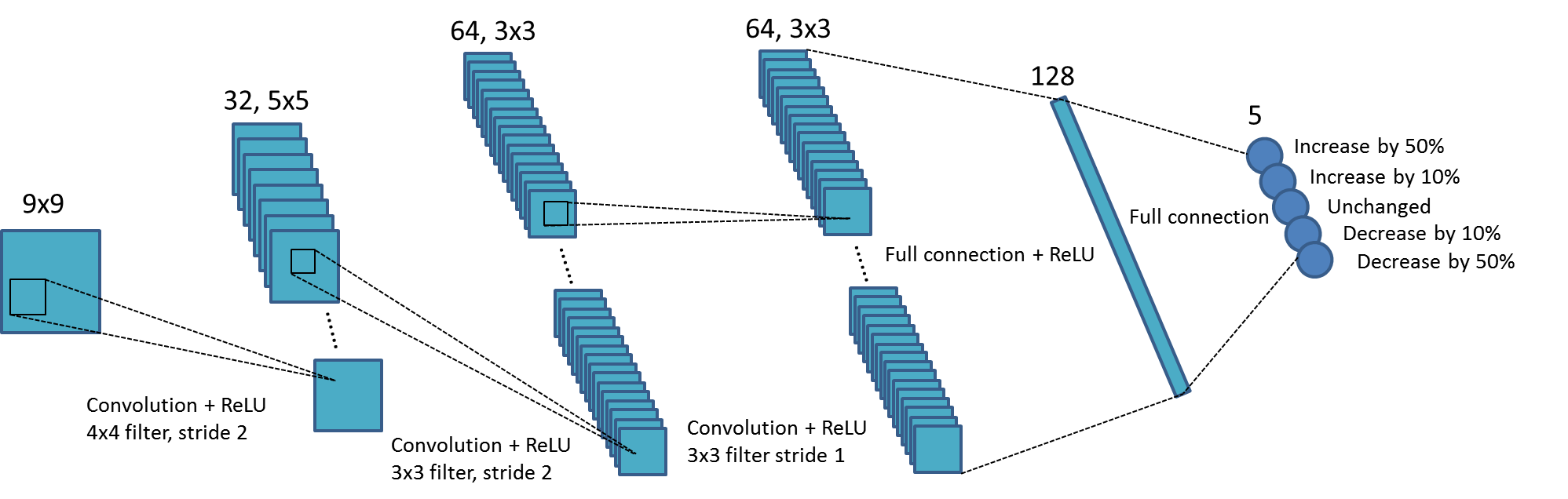}
  \caption{Network structure of PTPN. The input is a patch cropped out of the solution image $f^k$. The five outputs are directions and magnitudes of changing the parameter at the patch center pixel. Number of units and data sizes at each layer are specified at the top. Connection between subsequent layers are also presented.}
  \label{Fig:networkstructure}
\end{figure*}

Under this framework, we parameterize the value function $Q(s,a;W)$ using a convolutional neural network, where $W$ are network parameters. This network is referred as Parameter-Tuning Policy Network (PTPN) from here on. The structure of the network is depicted in Fig.~\ref{Fig:networkstructure}. $W$ will be determined through a reinforcement learning process, as will be described in the next section.

\subsection{PTPN training via deep reinforcement learning}

\subsubsection{General deep reinforcement learning idea}

One particular property of the $Q^*(s,a)$ function is the Bellman equation\cite{Bellman1964dynamic}:
\begin{equation}
	Q^*(s,a) = r + \gamma\max_{a'}~Q^*(s',a'),
\end{equation} 
where $s'$ is the state of the imaging system that follows $s$ after taking the action $a$. With this identity, for a function $Q(s,a)$, it is possible to define the loss function as the square of the deviation from this identity in order to quantify the deviation of $Q(s,a)$ from $Q^*(s,a)$. When the function $Q(s,a)$ is approximated with a network,  $Q(s,a)\approx Q(s,a;W)$, this loss function is $L(W)= [r + \gamma\max_{a'}Q(s',a';W) - Q(s,a;W)]^2$.

To determine $W$ through a reinforcement learning process, we introduce another variable $W'$ and hence define a target term $y = r + \gamma\max_{a'}Q(s',a';W')$. For a fixed $W'$, we consider the loss function 
\begin{equation}
	L(W)= [y - Q(s,a;W)]^2.
\label{Eq:Loss}
\end{equation}
Note that the $s'$ inside the target term is related to $s$ by the action $a$. At the end of the learning process, $W'$ and $W$ in  Eq.~(\ref{Eq:Loss}) should converge. This can be achieved by performing learning in a sequence of stages. In each stage, the parameter $W'$ is kept unchanged, whereas the parameter $W$ is optimized towards minimizing the loss function. At the end of each stage, $W'$ is updated to the optimized parameter $W$. 

Within a stage, since $W'$ is kept unchanged, the gradient of the loss function with respect to $W$ is simply
\begin{equation}
\begin{split}
	\frac{\partial L}{\partial W}=&[r+\gamma\max_{a'}Q(s',a';W')\\
	&\left.-Q(s,a;W)\right]\frac{\partial Q(s,a;W)}{\partial W}.
\end{split}
\label{Eq:grad}
\end{equation}
The last term $\partial Q(s,a;W)/\partial W$ can be computed via the standard back-propagation approach in a typical network training process. As in many other studies, we use stochastic gradient descent approach that computes the gradient and therefore updates the network parameter $W$ using a subset of training data randomly selected from the full training data set. $W$ is then updated as $W^{l+1}=W^l-\sigma\frac{\partial L}{\partial W}$, where $\sigma$ is learning rate and $l$ is the index of iteration. 

\subsubsection{Training PTPN}

We train the PTPN following the general idea outlined in the previous section. As such, we repeatedly perform image reconstruction using the ADMM algorithm in Fig.~\ref{Fig:ADMM}. At the step $k$, the solution image $f^k$ is observed. For each pixel $x$ we use a $\epsilon$-greedy to select an action to adjust the parameter value $\lambda(x)$. Specifically, with probability of $\epsilon$, we randomly select an action among all the possible choices with equal chances. Otherwise, we select the action $a^k$ that attains the highest output value of $Q(s,a;W)$ with the current image patch $s=S_{f^k}(x)$ as input; we choose $a =\arg\max_{a'}~Q(s,a;W)$. With the selected action, we update the parameter $\lambda^k(x)$ accordingly. After the parameters of all the pixels are updated, we perform image reconstruction one more time with $f^k$ as the initial guess, yielding an updated solution image $f^{k+1}$.

At this point, we randomly sample a number of $N_{samp}$ patches from the image  to generate training data for the PTPN. For each selected patch at location $x$, we gather the information of the solution image patches $S_{f^{k}}(x)$, $S_{f^{k+1}}(x)$,  the reward $r^k(x)$, as well as the action $a^k(x)$. The reward function at this patch is defined as 
\begin{equation}
r^k(x) = \frac{|S_{f^*}(x)|}{|S_{f^{k+1}}(x)-S_{f^*}(x)|}-\frac{|S_{f^*}(x)|}{|S_{f^{k}}(x)-S_{f^*}(x)|},
\end{equation}
where $f^*$ is the ground truth image. $|.|$ stands for the standard $L2$ norm of a vector.  We define this reward function to encourage image patch updates that are moving towards the ground truth image patch. The inverse function is utilized to amplify the change between $f^k$ and $f^{k+1}$: as the parameter is tuned through a sequence of steps, an additional step typically improves the image quality only slightly, and hence reduces the distance to the ground truth by a small amount. 

The collected information at different locations forms a set of data $\{s^k=S_{f^{k}}, r^k, a^k, s^{k+1}=S_{f^{k+1}}\}$. The data is then put into a pool of training data set. Finally, to train PTPN, a subset of the training data randomly selected from the pool are used to update parameter $W$ to minimize the loss function in Eq.~(\ref{Eq:Loss}) with gradient computed using Eq.~(\ref{Eq:grad}). This strategy is known as experience replay in the deep-Q learning regime, which is designed to overcome the problem that the training data generated in a sequential steps of actions are highly correlated\cite{Mnih:Arxiv:2013,Mnih:Nature:2015}. This process continues for a preset number of steps $N_{recon}$. Within this process, we update $W$ to $W'$, after every $N_{update}$ steps.

The training process described above is executed in multiple epochs. Each epoch contains the same training process on multiple data sets of different CT cases. The overall algorithm structure is summarized in Fig.~\ref{Fig:trainingAlgorithm}. 

\begin{figure}
\begin{algorithmic}[1]
\STATE{Initialize PTPN parameters $W$ and $W'$;}
\FOR {Epoch = $i,\ldots,N_{epoch}$}
	\FOR{Each training image $f^*$}
		\STATE{Initialize $\lambda^0(x)$;}
		\STATE{Reconstruct image $f^0$ under $\lambda^0(x)$ using ADMM;}	
		\FOR{$k = 0,\ldots,N_{recon}$}
			\STATE{Select an action $a^k$ for each image pixel $x$:}
			\STATE{\quad With probably $\epsilon$ choose $a^k$ randomly;}
			\STATE{\quad Otherwise $a^k=\arg\max_aQ(S_{f^k}(x),a;W)$;}
			\STATE{Compute $\lambda^k$ based on $a^k$ at each pixel;}
			\STATE{Reconstruct image $f^k+1$ under $\lambda^k$ using ADMM;}
			\STATE{Randomly sample $N_{samp}$ data for training:}
			\STATE{\quad Sample $N_{samp}$ pixels, for each pixel:}
			\STATE{\quad Get patches $S_{f^k}$, $S_{f^{k+1}}$, and $a^k$;}
			\STATE{\quad Compute reward $r^k$};
			\STATE{\quad Store $(S_{f^k}, S_{f^k}, r^k, a^k)$ in training data set;}
			\STATE{Train PTPN:}			
			\STATE{\quad Select $N_{train}$ data from training data set;}				\STATE{\quad Compute gradient using Eq.~(\ref{Eq:grad});}
			\STATE{\quad Update network parameter $W$;}
			\STATE{Set $W'=W$ every $N_{update}$ steps;}
		\ENDFOR
	\ENDFOR
\ENDFOR
\end{algorithmic}
\caption{Overall algorithm used to train the PTPN.}
\label{Fig:trainingAlgorithm}
\end{figure}

\subsection{Implementation details}
We implemented this algorithm using Python with TensorFlow. The computational platform is a desktop workstation with a Intel Xeon 3.5 GHz CPU processor, 8 GB memory and an Nvidia Quadro M4000 GPU card.

For the CT reconstruction part, we consider a fan-beam projection geometry with 180 projections equally spaced over a $2\pi$ angular range. The image has a resolution of $128\times 128$ pixels. A relatively low resolution is chosen due to computational concerns. The x-ray detector is of a line shape with $384$ elements covering a $40$ cm range. The source-to-isocenter distance is $100$ cm and the isocenter-to-detector distance is $50$ cm. The projection matrix $P$ is computed using the standard Siddon's algorithm \cite{Siddon:1985:MP}. We select six patient CT images at different anatomical sites including brain, lung, and abdomen as training images. Projection data is simply calculated as  $g=Pf^*+n$, where $f^*$ is the ground truth image and $n$ is a Gaussian noise signal with zero mean and variance determined by $Pf^*$ as in a previous study\cite{Wang:PMB:2008}. The averaged relative noise level is $3\%$. Values of relevant parameters used in training are summarized in Table~\ref{Table:parameter}.

\begin{table}
\centering
\caption{Relevant parameters used, when training the PTPN.}
\label{Table:parameter}
\begin{tabular}{c | c | l} 
 \hline 
   Parameter &  Value &  Comments \\ [0.5ex] 
 \hline\hline
 $\delta$ & $3\times 10^{-3}$ & Stopping criteria in ADMM \\
 $\gamma$ & 0.$99$ & Discount rate \\ 
 $\epsilon$ & $0.99\sim 0.1$ & Parameter of $\epsilon$-greedy approach \\
 $N_{epoch}$ & $100$ & Number of epochs \\
 $N_{samp}$ & $3200$ & \begin{tabular}{@{}c@{}}Number of sampled patches to add to \\ training data pool\end{tabular} \\
 $N_{recon}$ & $20$ & \begin{tabular}{@{}c@{}}Number of times to perform ADMM \\ reconstruction per epoch\end{tabular}  \\ 
 $\sigma$ & $0.001$ & Learning rate when updating $W$\\
 $N_{train}$ & 128 & Number of data for training each time \\
 $N_{update}$ & 300 & Number of steps to update $W'=W$ \\ 
 \hline
\end{tabular}
\end{table}

\section{Validation studies and results}

\subsection{Training process and trained PTPN}

\begin{figure}[b]
	\centering
  \includegraphics[width=0.4\textwidth]{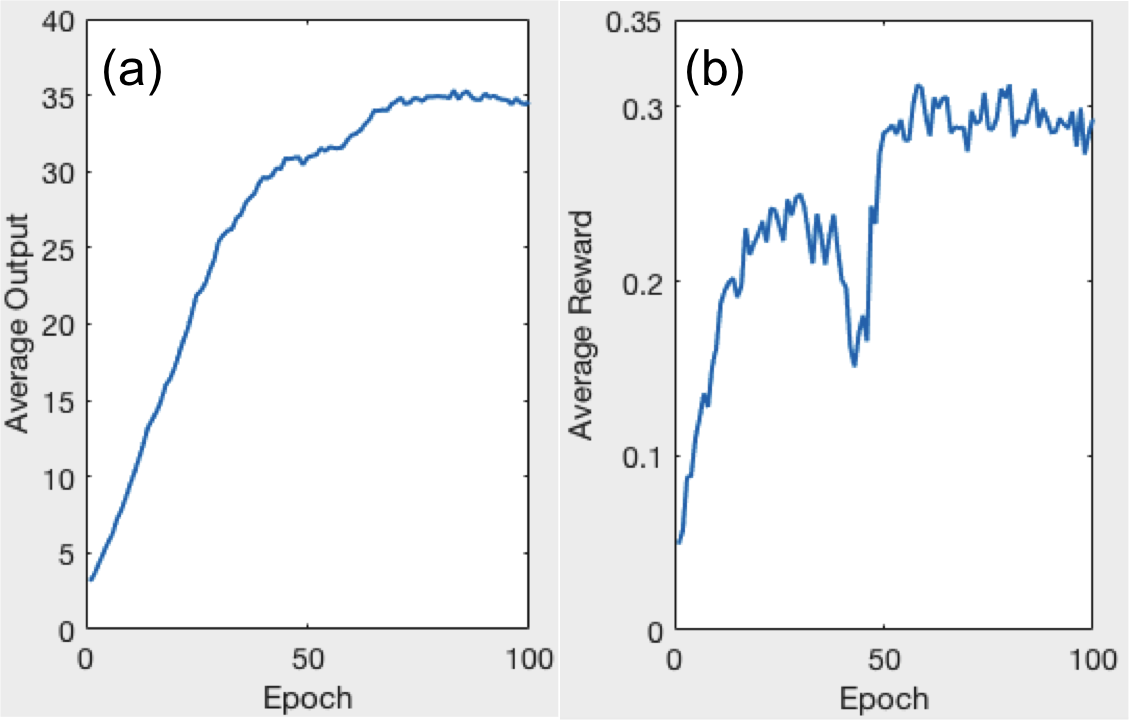}
  	\caption{Average output of PTPN (a) and reward (b) in each training epoch. }
  \label{Fig:training}
\end{figure}

During the training process, we monitor the quality of the trained PTPN shown in Fig.~\ref{Fig:training}. Both the average output of the PTPN and the reward follow an increasing trend although with some oscillations. This indicates that the PTPN is adjusted gradually in this reinforcement learning process towards predicting actions with high reward values. 

\subsection{Parameter tuning in CT reconstruction}

\subsubsection{CT reconstruction under PTPN guidance}
With the PTPN trained, we use it to guide parameter tuning in a CT reconstruction problem. As such, we select a ground truth CT image $f^*$ and generate the projection data with noise added. We first set the parameter arbitrarily to $\lambda^0(x)=0.005$, a constant value that is likely \emph{not} optimal. After that, we apply PTPN to guide parameter tuning as outlined in the first paragraph in Sec.II.B. The tuning process stops, when the relative difference between CT images in two successive reconstruction is less than $1\%$.

\begin{figure}
	\centering
  \includegraphics[width=0.4\textwidth]{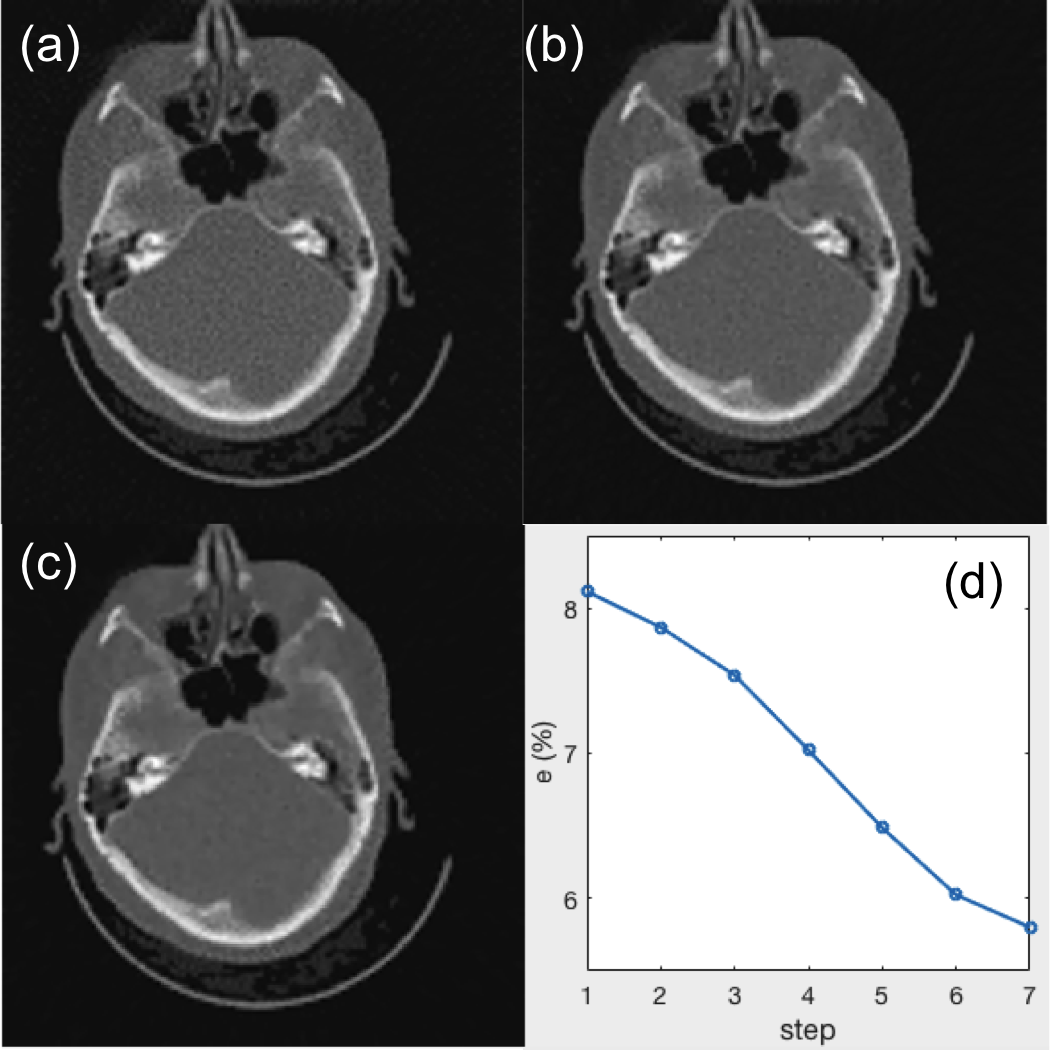}
  	\caption{(a)-(c) Reconstructed images at step 1, 4, and 7. (d) Error $e (\%)$ as a function of parameter tuning step. }
  \label{Fig:steps}
\end{figure}

To observe this process in detail, we select a test case that is \emph{not} used in training. Fig.~\ref{Fig:steps}(a)-(c) present reconstructed CT images at step 1, 4, and 7. It is clear that the image quality is improved with the parameter tuned. Quantitatively, we compute the relative error $e=|f-f^*|/|f^*|$ at different steps and plot it in Fig.~\ref{Fig:steps}(d). A monotonic decay trend is observed, indicating the effectiveness of parameter tuning.

\subsubsection{Reconstruction results}
\begin{figure}
	\centering
  	\includegraphics[width=0.5\textwidth]{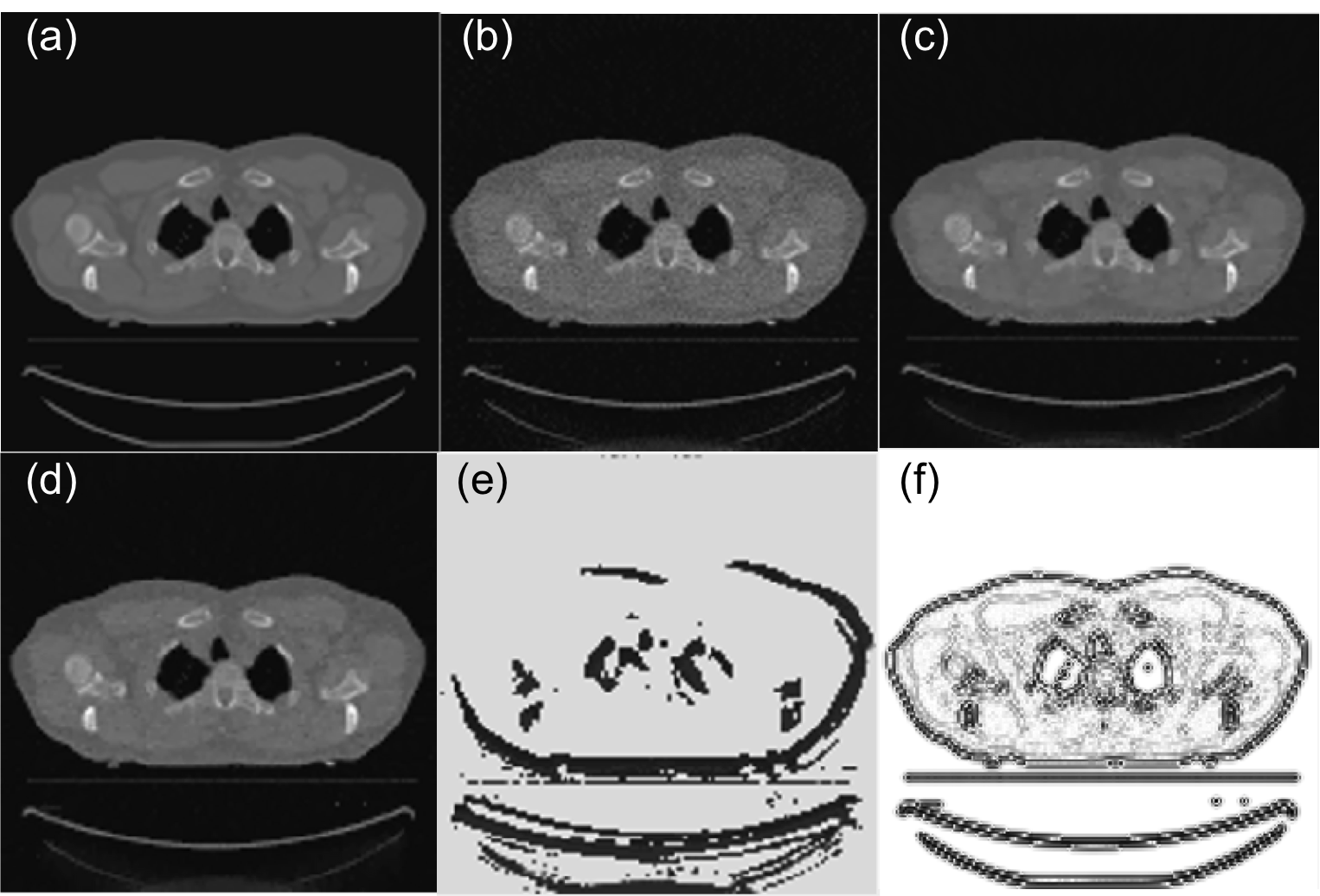}
  	\caption{(a) Ground truth CT image of a case that is used in training PTPN. (b) Image reconstructed with an arbitrarily selected parameter $\lambda(x) = 0.005$. (c) Image reconstructed after the parameter is tuned by PTPN. (d) Image reconstructed by manually tuned to $\lambda(x) = 0.05$. (e) Tuned parameter map $\lambda(x)$. (e) Optimal parameter map $\lambda^*(x)$.  }
  \label{Fig:result_train}
\end{figure}

\begin{figure}[b]
	\centering
  	\includegraphics[width=0.5\textwidth]{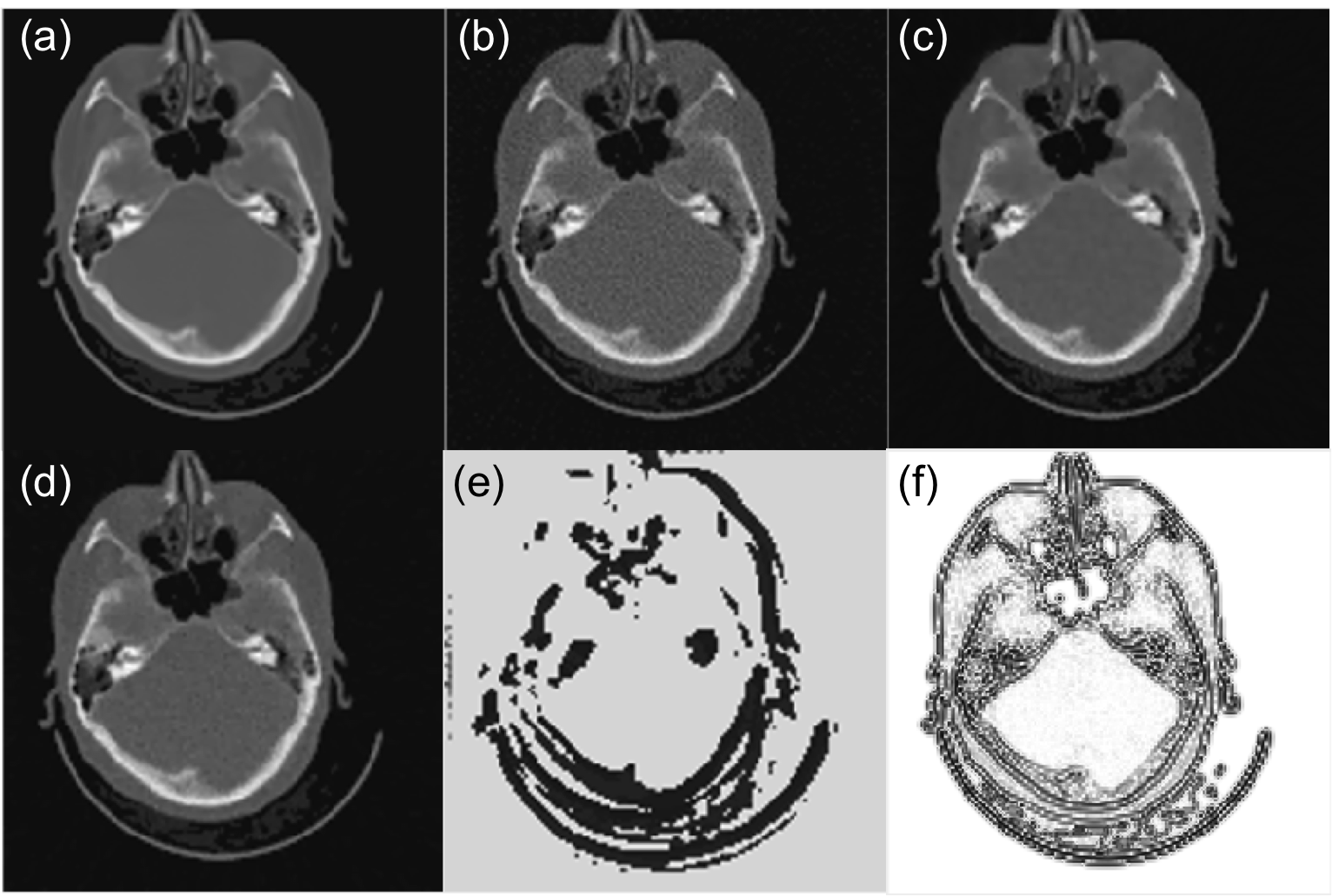}
  	\caption{(a) Ground truth CT image of a case that is not used in training PTPN. (b) Image reconstructed with an arbitrarily selected parameter $\lambda(x) =0.005$. (c) Image reconstructed after the parameter is tuned by PTPN. (d) Image reconstructed by manually tuned to $\lambda(x) = 0.12$. (e) Tuned parameter map $\lambda(x)$. (e) Optimal parameter map $\lambda^*(x)$.  }
  \label{Fig:result_test}
\end{figure}

Fig.~\ref{Fig:result_train}  is a case that is used in training, whereas Fig.~\ref{Fig:result_test} is the same one in Fig.~\ref{Fig:steps}, which is not included in training. Since we arbitrarily set initial values of $\lambda(x)$, which is too small in these two cases, the resulting images contain a lot of noise (Fig.~\ref{Fig:result_train}(b) and \ref{Fig:result_test}(b)). After the parameter $\lambda(x)$ is tuned by PTPN, the image quality in both cases is substantially improved (Fig.~\ref{Fig:result_train}(c) and \ref{Fig:result_test}(c)). 

We compare the results with those under manually tuned parameters. Since it is impractical for one to adjust the parameter for each individual pixel, we consider a special context that the parameter is a constant throughout the image and we manually adjust this parameter value for the best image quality. The appropriate parameter values are $\lambda(x)=0.05$ for Fig.~\ref{Fig:result_train} and $\lambda(x) = 0.12$ for Fig.~\ref{Fig:result_test}. Fig.~\ref{Fig:result_train}(d) and \ref{Fig:result_test}(d) depict images reconstructed under these parameters in the two cases, respectively. It is found that the images still contain a certain amount of noise and the quality is inferior to those with parameters tuned by PTPN. 

As for the parameter maps tuned by the PTPN shown in Fig.~\ref{Fig:result_train}(e) and \ref{Fig:result_test}(e), it is observed that PTPN deliberately reduces parameter values most around image edges. This is understandable. Reducing parameters at those pixels decreases the amount of regularization in those areas, which is beneficial in terms of preserving image edges. 

Interestingly, for the simple problem in Eq.~(\ref{Eq:CTrecon}), it is possible to derive the optimal parameter map $\lambda^*(x)$. As such, let us take the gradient of the objective function and set it to zero at $f=f^*$: $\left.P^T(Pf-g)-\lambda\nabla\cdot\left(\frac{\nabla f}{|\nabla f|}\right)\right|_{f=f^*}=0$.
This implies that the optimal parameter map is 
\begin{equation}
	\lambda^*(x)=\frac{P^T(Pf^*-g)}{\nabla\cdot\left(\frac{\nabla f^*}{|\nabla f^*|}\right)}.
\end{equation}  
The numerator in this expression is more or less an image of noise that is obtained by back-projecting the residual error in the projection domain to the image domain. Here, we neglect the image structure of the noise and plot the image $1/\nabla\cdot\left(\frac{\nabla f^*}{|\nabla f^*|}\right)$ in Fig.~\ref{Fig:result_train}(f) and \ref{Fig:result_test}(f) for the two cases, respectively. The images shows that $\lambda^*(x)$ is small along the image edges. Comparing subfigures (e) and (f) in Fig.~\ref{Fig:result_train} and Fig.~\ref{Fig:result_test}, the similarity between corresponding pair of images implies that PTPN can intelligently adjust $\lambda(x)$ towards the optimal parameter maps. Note that this intelligence is purely developed by the PTPN itself through the reinforcement learning process. Except providing rewards for an action, we do not explicitly give any information regarding how to tune the parameters. 

Quantitatively, we evaluate the image quality using relative error $e$ and Peak Signal-to-Noise Ratio (PSNR). Table~\ref{Table:metrics} summarizes the results in six training and six testing cases. These cases are CT images at different anatomical sites. In each case, we present the metrics for the image under manually tuned parameters, under an arbitrarily set initial parameter, and under parameter tuned by PTPN.  For all the training cases, the images under PTPN-tuned parameters achieve the smallest error and the highest PSNR, indicating the satisfactory quality of the trained PTPN. Among the six testing cases, the PTPN-tuned parameter yields the smallest errors and the highest PSNRs in five cases ($\#$1-4, 6). For the case $\#$5, the difference between manually tuned and PTPN-tuned results is small.

\begin{table*}
\centering
\caption{Relative error and PSNR of image reconstruction results. Boldface numbers indicate the best result in each case. }
\label{Table:metrics}
\begin{tabular}{c|c||c|c|c||c|c|c}
            \hline
            \multicolumn{2}{c||}{Case} & $e_{Manual} (\%)$ & $e_{Initial} (\%)$ & $e_{Tuned} (\%)$ & $PSNR_{Manual}$ (dB) & $PSNR_{Initial}$ (dB)& $PSNR_{Tuned}$ (dB)\\ [0.5ex]
            \hline\hline
             \multirow{6}{*}{Training} & 1 & 4.47 & 7.50 & {\bf 4.21} & 39.13 & 34.65 & {\bf 39.67}\\
              & 2 & 4.69 & 7.72 & {\bf 4.56}& 38.67 & 34.33 & {\bf 38.90}\\
              & 3 & 10.77 & 11.89 & {\bf 10.38}& 31.57 & 30.70 & {\bf 31.89}\\
              & 4 & 12.92 & 13.61 & {\bf 12.54}& 29.56 & 29.11 & {\bf 29.82}\\
              & 5 & 3.68 & 6.83 & {\bf 3.62}& 40.04 & 35.07 & {\bf 40.59}\\
              & 6 & 3.82 & 6.78 & {\bf 3.55}& 41.09 & 36.11 & {\bf 41.74}\\
            \hline
            \multirow{6}{*}{Testing} & 1 & 4.35 & 6.93 & {\bf 4.24}& 44.28 & 40.22 & {\bf 44.50} \\
              & 2 & 12.17 & 12.35 & {\bf 12.13}& 29.45 & 29.32 & {\bf 29.48}\\
              & 3 & 10.30 & 11.49 & {\bf 8.48}& 32.14 & 31.19 & {\bf 33.83}\\
              & 4 & 5.42 & 8.17 & {\bf 5.32}& 36.67 & 33.10 & {\bf 36.89}\\
              & 5 & {\bf 4.62} & 7.19 & 4.95& {\bf 36.91} & 33.07 &  36.31\\
              & 6 & 8.56 & 10.29 & {\bf 7.56}& 31.69 & 30.09 & {\bf 32.78}\\
            \hline
        \end{tabular}
\end{table*}

\subsubsection{Application to other cases}

\begin{figure}[t]
	\centering
  	\includegraphics[width=0.5\textwidth]{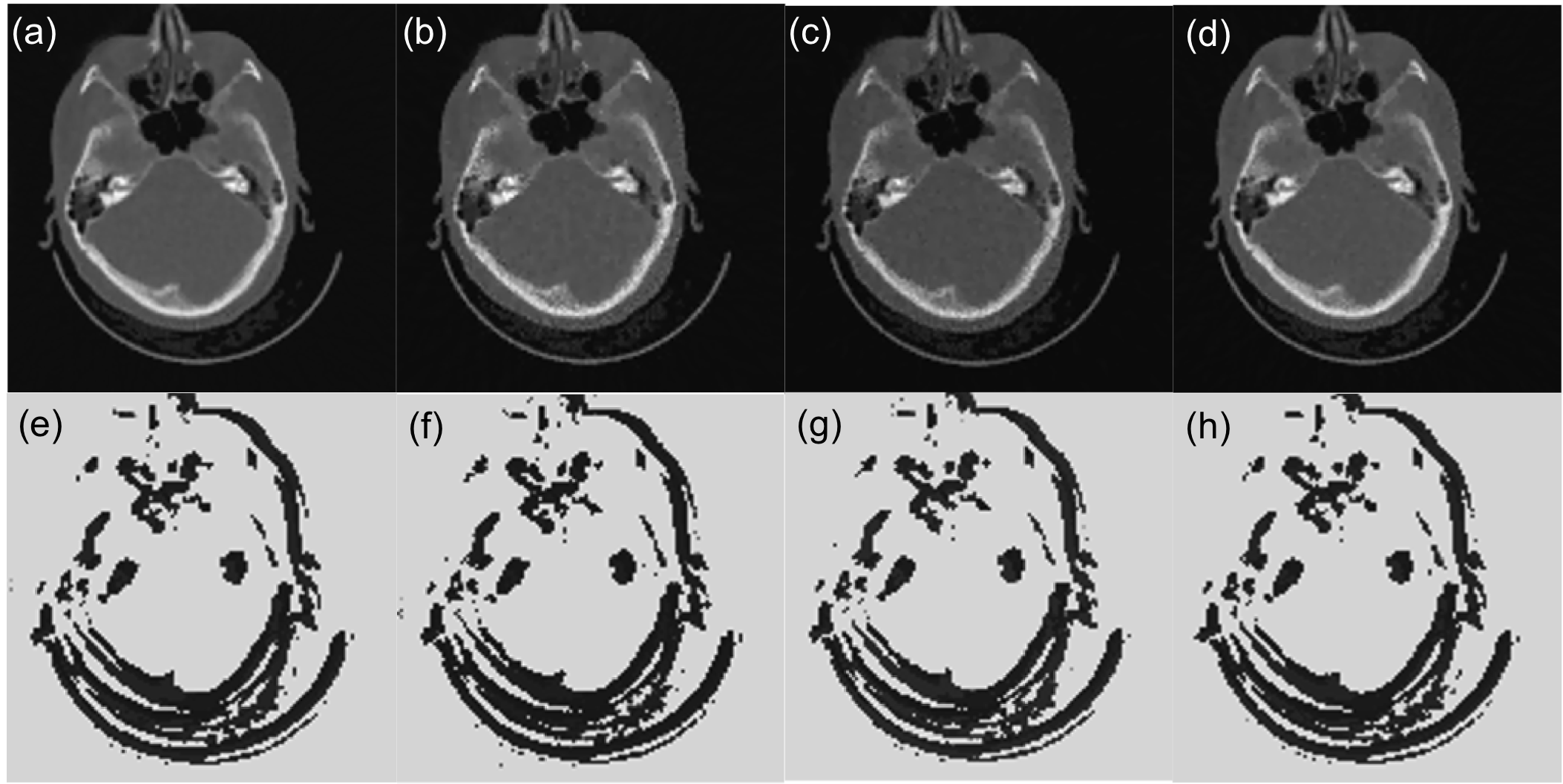}
  	\caption{(a)-(b) The results under tuned parameter for a case with $2\%$ and $5\%$ noise in the projection data. (c) The result with $90$ projections. (c) The result with isocenter-to-detector distance changed to $25$ cm. Figures in the bottom row are tuned parameter maps for corresponding figures in the top row.   }
  \label{Fig:otherCases}
\end{figure}

The PTPN determines the way of parameter tuning based on observed image patch. It is expected that the trained PTPN is also applicable to image reconstruction under settings that are different from that in training. To demonstrate this fact, we also applied PTPN to image reconstruction in cases with different number of projections, noise levels, and projection geometry. Fig.~\ref{Fig:otherCases}(a) and (b) are the same case as in Fig.~\ref{Fig:result_test} but with $2\%$ and $5\%$ noise in the projection data, different from the noise level of $3\%$ in training. Fig.~\ref{Fig:otherCases}(c) is the case with only $90$ projections. In Fig.~\ref{Fig:otherCases}(d) we change the isocenter-to-detector distance to $25$ cm. In all the cases, PTPN is able to adjust parameters to yield images with satisfactory quality. The resulting parameter maps in Fig.~\ref{Fig:otherCases}(e)-(h) are all similar to the ground truth shown in Fig.~\ref{Fig:result_test}(f). 

\section{Discussions}

\emph{Relation to other works.} The power of deep learning in medical image processing has been clearly demonstrated in a spectrum of problems. Among these studies, most used supervised training to determine parameters inside a network in order to establish a map between input and output images. In \cite{Chen:BOE:2017,Chen:TMI:2017}, a network was set up to map a noise-contaminated CT image acquired at a low-dose level to the clean image. In \cite{Han:CoRR:2016}, deep residual learning was employed to map a CT image with streak artifacts caused by undersampling to the artifact image, which was further subtracted from the original image to eliminate the artifact. In an study \cite{Yang:ANIPS:2016} that viewed the iterative image reconstruction process as a data flow in a network under the ADMM algorithm, the supervised learning process enabled discovery of the algorithm parameters, such as image filters and threshold values. Comparing to these novel works, our study is different in twofold. First, the purpose of using deep learning is different. Instead of trying to \emph{predict the underlying true solution or image artifacts}, the purpose of setting up a PTPN is to \emph{predict a dynamic policy} applicable to the image reconstruction problem in Eq.~(\ref{Eq:CTrecon}). Under the guidance of this policy, the output image of the reconstruction algorithm is directed towards a satisfactory quality. Second, the method to train our network is also different from the supervised training in previous works. Instead of using labeled training pairs in a supervised training fashion, we employed the reinforcement learning strategy. This strategy let the algorithm to play by itself and get rewards based on the image and the selected action. Through the training process, the PTPN spontaneously discovered the appropriate strategy for an input system state. This was the process in which intelligence is generated. 

The CT reconstruction problem with pixel-wise regularization has been investigated in previous studies\cite{Guo:SPIE:2010,Tian:PMB:2011}. It was proposed to perform a sequence of reconstructions with parameters adjusted based on the reconstructed images. The motivation was to detect image edges and to tune down the regularization weights for the purpose of edge preservation. As opposed to designing this explicit rule of parameter tuning, this study discovered the rule via the reinforcement learning process. It is interesting to observe that intelligence can be correctly generated, which coincides with previous human knowledge.

\emph{Necessity of deep reinforcement learning.} What is ultimately learned by the PTPN is evaluation of the image quality and the link to parameter tuning. With this in mind, one may argue that the complex reinforcement learning technique is probably unnecessary, as one can simply perform supervised learning by using a sizable data set containing paired data of image patches and corresponding ways of parameter tuning. We agree with this statement to a certain extent, but still think our study is meaningful. For this CT reconstruction problem, it is straightforward to generate labeled training pairs (image patch and direction of parameter tuning) to allow supervised training. Yet if we would like to label an image patch with not only the direction of parameter tuning, i.e. increase or decrease, but also with the amount of parameter change, i.e. $50\%$ or $20\%$ as in our example, it becomes quite difficult to generate training data. Hence, the advantage of reinforcement learning is to automatically learn a more comprehensive policy. Beyond the problem of CT reconstruction, it may not be easy to generate labeled training pairs in many optimization-based inverse problems. However, since very often one has a good sense of judging the output results, it is still relatively easy to quantify the result quality via a reward function. This allows the use of reinforcement learning to establish the policy in those problems for which labeled training pairs are hard to get.  

\emph{Relevance to other problems.} This study uses an optimization-based iterative CT reconstruction problem as an example to show that it is possible to achieve intelligent parameter tuning via a deep learning approach. With the rapid growth of deep learning techniques in CT reconstruction area, the impact of this study may diminish. However, we think studying the general task of parameter tuning is still of significance and deep learning opens a new window to tackle this problem. First, parameter tuning is not a problem unique to the CT reconstruction regime, but generally existing in many areas. Even beyond the scope of image processing, many other decision making problems in medicine can be solved in an optimization approach, for which parameter tuning is an indispensable task. One notable example is treatment planning in cancer radiation therapy\cite{bortfeld2006imrt}. Even with a modern treatment planning system to solve the underlying optimization problem, a hospital still needs to hire a number of dosimetrists to manually tune the parameters in order to generate plans meeting clinician's requirements. This fact clearly highlights the needs for and potential benefits of an intelligent parameter tuning system. Second, even for the deep learning technique itself, the training stage has a number of parameters to be tuned by the researcher to achieve the best performance. These parameters include, but are not limited to, learning rate, number of epochs, size and number of filters, etc. It would be an interesting and important step to develop a parameter tuning system to handle the adjustment of these parameters. Meanwhile, we have to admit that solving the parameter tuning problem in the area beyond the simple example of CT reconstruction is apparently much more challenging. We hope our study here can shed some light in this direction and trigger deeper investigations in future. 

\emph{Limitations and future directions.} This study has the following limitations. First, due to limitation on computational power, we only considered images with a relatively low resolution in a small number of cases. It is our plan to extend the studies to high-resolution images that are of more clinical relevance. We will also use more cases for training and testing to yield a more robust PTPN. The second limitation of this study is that PTPN has to wait for the ADMM iterative process to finish, before it can adjust parameters. Although the image quality resulting from this this approach is acceptable, waiting for the ADMM to finish reduces the overall workflow efficiency. This can be potentially improved by using another reconstruction algorithm with a higher convergence rate. Another possible way of acceleration is to predict the converged CT image at an early step of the iterative ADMM reconstruction, for instance using a deep learning approach\cite{Cheng:Fully3d:2017}. Third, PTPN lays a general framework on developing strategy to improve image quality. The current setup in Eq.~(\ref{Eq:CTrecon}) limits possible policy to the five options of modifying the regularization parameters. However, in general, it is possible to include policy that act more directly on the images, such as reducing noise and artifacts, etc. It is noted that deep-learning has achieved tremendously in each of these CT image enhancement problems\cite{Han:CoRR:2016,Kang:CoRR:2016,Li:Fully3d:2017}. Using them under the guidance of a policy network is expect to yield a complete and comprehensive image reconstruction system that can automatically handle various of data contamination. 

\section{Conclusion}

In this paper, we have aimed ourselves at shedding some lights to the task of automatic parameter tuning in an optimization problem, which is a typical task in a number of image-processing, or non-image-processing problems. The significance of this study is underscored by the fact that the solution quality is critically determined by the parameter values, and yet there is no satisfactory way of automatically adjusting parameters. We proposed to solve this problem by constructing a policy network, which can be trained to guide parameter tuning. We demonstrated our idea in an example problem of optimization-based iterative CT reconstruction with a pixel-wise TV regularization term. We configured a PTPN to map a CT image patch to the direction and magnitude of tuning the parameter at the patch center. PTPN was trained via an end-to-end reinforcement learning procedure. A series tests demonstrated that the trained PTPN is able to intelligently determine the way of parameter adjustment. Under the guidance of PTPN, the reconstructed CT images achieved image quality similar or better than that under manually tuned parameters.


%



\section*{Acknowledgment}

The authors would like to thank funding support from NIH grant 1R21EB021545.


\bibliographystyle{IEEEtran}
\bibliography{reference}

\end{document}